\documentclass [twocolumn, aps, prb, showpacs] {revtex4}
\usepackage{graphicx}
\usepackage[dvips]{color}
\input{epsf}

\def\ET{{$\kappa$-(ET)$_2$Cu$_2$(CN)$_3$}}

\def\etal{~\textit{et~al.}} 
\def\ra{\rangle} 
\def\la{\langle} 
\def\up{\uparrow}
\def\dn{\downarrow}
\begin{document}

\title {Impurities in triangular lattice spin 1/2 antiferromagnet}

\author{Karol Gregor and Olexei I. Motrunich}
\affiliation{Department of Physics, California Institute of
Technology, Pasadena, CA 91125}
\date{\today}

\begin{abstract}
We study effects of nonmagnetic impurities in a spin-1/2 frustrated
triangular antiferromagnet with the aim of understanding the observed
broadening of $^{13}$C NMR lines in the organic spin liquid material
$\kappa$-(ET)$_2$Cu$_2$(CN)$_3$.
For high temperatures down to $J/3$, we calculate local susceptibility
near a nonmagnetic impurity and near a grain boundary for the nearest
neighbor Heisenberg model in high temperature series expansion.
We find that the local susceptibility decays to the uniform one in
few lattice spacings, and for a low density of impurities we would
not be able to explain the line broadening present in the experiments
already at elevated temperatures.
At low temperatures, we assume a gapless spin liquid with a
Fermi surface of spinons.  We calculate the local susceptibility in the
mean field and also go beyond the mean field by Gutzwiller projection.
The zero temperature local susceptibility decays as a power law and
oscillates at $2 k_F$.  As in the high temperature analysis we find that
a low density of impurities is not able to explain the observed
broadening of the lines.  We are thus led to conclude that there is more
disorder in the system.  We find that a large density of point-like
disorder gives broadening that is consistent with the experiment down to
about $5$K, but that below this temperature additional mechanism is
likely needed.
\end{abstract}

\maketitle

\section{Introduction}

Spin liquid phases are some of the most interesting phases known to
exist theoretically.  However they are hard to achieve experimentally
because interactions usually favor ordered phases.  To achieve spin
liquid we have to frustrate these interactions.  Triangular lattice
provides a natural way to do this.  For the nearest neighbor
antiferromagnetic Heisenberg model the frustration is not strong enough
and the ground state is ordered.  However, the order is weak, and it is
likely that in the presence of appropriate additional interactions,
spin liquid phases arise.

This work is motivated by the layered organic compound \ET.\cite{Shimizu03, Kurosaki, Kawamoto04, Shimizu06, Kawamoto06, McKenzie}
It contains ET molecules that pair up, each pair lies on sites
of triangular lattice and has one electron less then the full filling.
The material at ambient pressure is an insulator.
Thus it is effectively a spin 1/2 antiferromagnet on the triangular
lattice.  While the Heisenberg exchange $J \approx 250$~K, the system
shows no signs of ordering down to $32$~mK making it a good candidate
for the spin liquid.
There are likely additional interactions among spins, especially ring
exchanges that are thought to be responsible for driving the system
into the spin liquid.\cite{LiMing, ringxch, SSLee, Nave}
What makes the appearance of such interactions natural is that under
moderate pressure the \ET\ undergoes a transition to a superconductor
at low temperature (and a metal at higher temperature), so there
are significant virtual charge fluctuations present already in the
insulator at ambient pressure.\cite{Shimizu03, Kurosaki}
Crudely, we can think of the system as a half-filled Hubbard model
close to the Mott insulator - metal transition, and we can estimate
that the ring exchange interactions in the effective spin model are
strong, enough to destroy the magnetic order.
\cite{ringxch, Zheng, LiMing}
An alternative explanation of the insulator in terms of inhomogeneous
electron localization has also been suggested.
\cite{Kawamoto04, Kawamoto06}

The spin liquid phase remains enigmatic.  Thermodynamic measurements
show many gapless excitations in this charge insulator -- at least
as many as in a metal.  One appealing proposal that captures some of the
observed phenomenology is a state with spinon Fermi surface.
\cite{ringxch, SSLee, Nave}  Other scenarios have also been suggested,
\cite{Wang, Galitski, Qi, Senthil}  particularly with the view
towards low temperatures.

We are specifically interested in the $^{13}$C NMR measurements of the
Knight shifts,\cite{Kawamoto04, Shimizu06, Kawamoto06}
and what we can learn from these for the material and the spin liquid.
The measurements effectively give a histogram of local magnetic
susceptibility and are therefore a good probe of the magnetic properties.
The experiment shows strong broadening of such histogram as one lowers
the temperature.  The width of the peak broadens by about a
factor of 40 as the temperature is lowered from 250~K to roughly 1~K and
saturates as the temperature is lowered further.
The distribution of local susceptibilities can be produced by disorder,
since the susceptibility can have various values as a function of
distance from say an impurity.  It is hard to imagine other mechanism
producing a distribution (except spin glass, but no such behavior is
observed).  Therefore in this paper we investigate the effects of
disorder on the spin system.

Unfortunately, not much is known about the impurities and their role
in the insulating \ET\ at ambient pressure.
At high pressure of 0.8~GPa, the material is a relatively clean metal
with $k_F l \gtrsim 50 - 100$ and observable Shubnikov-deHaas
oscillations.
It is believed that the Cu$^{2+}$ impurity concentration is very
low,\cite{Shimizu03} less than 0.01\%.  There are quite possibly
additional sources of disorder such as different local environments
coming from different conformations of the ET molecules,
\cite{Soto, Miyagawa, Wolter, Maksimuk}
or from disorder in the insulating anion layers such as the
disordered (CN)$^{-}$ group.\cite{Geiser, Komatsu, Emge, Kawamoto06}
Analyzing the NMR experiments can then provide some understanding
of the disorder, its strength, and role in the insulator phase.
Taking up the Mott insulator picture as one viable candidate, where the
insulator is primarily driven by electron-electron interactions,
we set out to study models of non-magnetic disorder in a spin-1/2 system
on the triangular lattice.  We study progressively different kinds of
disorder and analyze what each predicts about the local susceptibilities
in turn.

This paper consists of two separate approaches: one, the high
temperature series expansion, and the other, low temperature analysis
assuming the system forms a spin liquid.  High temperature series
expansion is rather restricted by the range of temperatures and the types
of models it can study, but for those models it gives exact results.
We are able to calculate local susceptibilities of the nearest neighbor
Heisenberg model on an arbitrary graph where all exchange couplings
are the same.  In particular, we can study triangular lattice with a
missing site, with a boundary, or in the presence of a finite density
of missing sites.  The missing sites are models of nonmagnetic
impurities, while the boundary is a model of grain boundary.
We can go down to temperature of roughly $J/3$.  In this range,
the experiments already see broadening of the Knight shift distribution
by about a factor of two and thus we can compare the calculated results
to the experiments.
Of course, the real material has interactions beyond the nearest neighbor
since it is a spin liquid, but rough estimate can be made, especially
given that multi-spin exchanges are less important at high temperatures.

At low temperatures we assume phenomenologically the system forms a
spin liquid with Fermi sea of spinons (stabilized by additional
interactions).  We first analyze this in mean field where it
reduces to free fermions hopping on the triangular lattice.
The spin liquid can naturally accommodate nonmagnetic disorder in the
spin model by the corresponding changes in the spinon hopping amplitudes.
The full theory also contains a dynamical U(1) gauge field,
\cite{SSLee, LeeNagaosaWen, Ioffe, LeeNagaosa, Polchinski, Altshuler}
but this is hard to analyze directly.
Instead, to go beyond mean field we study wavefunctions obtained by
Gutzwiller projection of the mean field states.  In one dimension, this
can capture the full theory, while in two dimensions this is only
an approximation but a reasonable one and dealing directly with
physical spins.

Overall, we find that the local susceptibility decays rather quickly
near an impurity and at small impurity densities such as $0.01\%$ of Cu
our results are very far from explaining the experimental data --
they would produce very sharp histograms as most of the sites are
in the bulk.
In the spin liquid phase, the local susceptibility has an oscillatory
$2k_F$ component that decays with a power law envelope away from defects,
so the impurities can be felt at larger distances, but the overall
amplitude that we find is still small.

We then studied the system at high temperature near a boundary and
in the presence of a larger density of missing sites.
We also studied the system at low temperature in the mean field near a
boundary, in the presence of a larger density of missing sites,
and in the presence of random disorder on bonds which is either
uniform or localized at a fraction of bonds.
From this analysis it appears that the most likely scenario is the case
of relatively large density of point-like disorder, where the linewidths
broaden with lowering the temperature until the correlation
length becomes comparable to the typical distance between defects.
A puzzling feature is that with such fixed disorder we cannot
reproduce the observed strong temperature dependence of the NMR lines
as the temperatures are lowered further.
On the other hand, such models in the metallic phase
where we considered electrons with random on-site potentials match
reasonably with the experiments under pressure, where the NMR linewidths
remain unchanged with temperature.\cite{Kawamoto06}
It could be also that the effective strength of disorder increases as the
temperature is lowered in the insulator, perhaps because of the
vanishing screening of charged impurities.  Better understanding
of the disorder in the \ET\ system is clearly needed.

A new triangular lattice spin liquid material
EtMe$_3$Sb[Pd(dmit)$_2$]$_2$, Ref.~\onlinecite{Itou},
has rather similar phenomenology to that of the \ET\ and also appears to
have significant NMR line broadening,
which may thus be a common feature of gapless spin liquids.
It would be interesting to compare both systems more.

Finally we would like to mention that we have performed similar analysis
on Kagome antiferromagnet addressing the NMR line broadening in the
candidate spin liquid material ZnCu$_3$(OH)$_6$Cl$_2$.
\cite{Kagome_SL, Kagome_NMR, Kagome_Zn}
There, the disorder is relatively well understood experimentally
and is estimated to be about $5\%$ vacancies.
Our calculations\cite{kagome_w_vacancies} in this case compare
sensibly with the experiment.

\section{Summary of the Experimental Inhomogeneous Line Broadening}

In what follows, we calculate local susceptibility in spin models
with nonmagnetic disorder.  To set the stage, we summarize the main
experimental findings in the form convenient for judging theoretical
results.
A direct comparison with the experiments is to look at the width
of the local susceptibility histograms relative to the average
susceptibility.  In the model calculations the bulk susceptibility is
roughly the location of the histogram peak, while in the experimental
plots we should also be aware of the chemical shifts.
The referencing to the average susceptibility is justified since in the
\ET\ this remains roughly unchanged around
$\bar{\chi} = 5 \cdot 10^{-4}$~emu/mol in a wide temperature range
between 300~K and 30~K and then decreases somewhat to a value
around $\bar{\chi} = 3 \cdot 10^{-4}$~emu/mol at 1~K.
Also, the bulk values can be quantitatively reproduced by
suitable choices of the model parameters such as $J$ in the
high-temperature series study\cite{Shimizu03, Zheng} or the spinon
hopping amplitude in the spin liquid model (see Sec.~\ref{sec:SL}).

Refs.~\onlinecite{Kawamoto04, Shimizu06, Kawamoto06} show the NMR lines
plotted versus shifts from tetramethylsilane (TMS),
in parts per million (ppm).  For the more strongly coupled $^{13}$C
whose hyperfine coupling constant is $0.21$~Tesla/($\mu_B$~dimer), the
susceptibility of $\bar{\chi} = 5 \cdot 10^{-4}$~emu/mol corresponds to
$\delta B/B = 1.8 \cdot 10^{-4} = 180$~ppm Knight shift.
Thus a 20~ppm shift corresponds to roughly a 10\% relative change in the
susceptibility at higher temperatures and a somewhat larger relative
change at lower temperatures.  Reading from the experiments,
the full width at half maximum (FWHM) of the line measured in such
relative terms increases from about 10\% at 300~K to about
20-30\% at 50~K to about 50-60\% at 10~K, and then increases steeply
to about 300\% at 1~K and saturates around this value at still lower
temperatures.
One can appreciate the dramatic broadening directly from the line shapes
in Refs.~\onlinecite{Kawamoto04, Shimizu06, Kawamoto06}, where the
distance of this line from the origin sets a natural scale since the
Knight shift and the chemical shift are comparable.
(Note that to see the inhomogeneous Knight shifts over the intrinsic
linewidths the experiments are done in large fields of order 8~Tesla;
this may be inducing more significant ground state modifications, while
in this work we focus on the linear response susceptibilities and their
distributions.)

We note that the broadening first sets in gradually coming from
high temperatures.  The line is already 20-30\% broad at
$T = 50~{\rm K} \sim J/5$ , which we can hope to understand
quantitatively using reliable high-temperature series approach.
It further broadens by about a factor of two to three in the region
50~K to 5-10~K where we expect the spin liquid approach to become
applicable.  We now present these two studies.

\section{High Temperature Series Expansion}

We consider spin $1/2$ nearest neighbor anti-ferromagnetic Heisenberg
model on triangular lattice.  Local spin susceptibility at site $i$ is
given by
\begin{equation}
\chi_{\rm loc}(i) =
\frac{(g\mu_B)^2 \la S_i^z S_{\rm tot}^z \ra}{k_B T} ~,
\end{equation}
where $S_{\rm tot}^z = \sum_j S_j^z$.  We calculate $\chi$'s in the
high temperature series expansions in the presence of nonmagnetic
impurities treated as missing sites (vacancies), and also near an
open boundary which is a model for a grain boundary.

The expansion is performed to the 12-th order in $J$ (or 13-th order
in $1/T$) using the linked cluster expansion.\cite{OitmaaBook}
The outline of the procedure is as follows.  We generate all abstract
graphs up to desired size.  Then we generate all subgraphs of these
graphs, keeping track of the location of each subgraph in the graph.

We calculate the local susceptibility for each graph at each point of
the graph.  Then we subtract all the subgraphs of each graph as needed
in the linked cluster expansion to get the contribution this graph would
give when embedded into lattice.  The local susceptibility on any
lattice can be calculated by creating all possible embeddings of all
graphs and adding their contributions at every site.
In this general formulation the lattice does no need to be regular,
it can be any connected graph.  In particular this procedure applies to
the triangular lattice with a missing site, with a finite density of
missing sites, or with a boundary.  In practice, for the single impurity
case, we consider all the graphs containing the impurity and subtract
their contribution from the uniform susceptibility.  Similar procedure
is used in the case with the boundary. In the end, we obtain exact
$1/T$ series expansion for the system with such disorder.\cite{kagome_w_vacancies}

After obtaining the series, we extend it beyond the radius of
convergence using the method of Pade approximants.  We use
[5,6], [5,7], [6,6], [6,7] and expand in variable $1/(T+\alpha)$ where
$\alpha$ is usually 0.08 as in Ref.~\onlinecite{Elstner93}.
Depending on $\alpha$ one might get a pole in the expression,
and hence divergence in susceptibility even at relatively large
temperature.  This usually happens say in
one of the approximants while the others still overlap. At low enough
temperatures they start diverging and we take that as a point where
the approximation stops being valid.  Different values of $\alpha$ are
tried, and sometimes it is possible to tune to a value where all the
curves overlap completely to a much lower temperature, but that is a
pathology, probably indicating that the polynomials are all the same.
For other values of $\alpha$ the curves usually start to diverge from
each other at around the same temperature.

\subsection{Point Impurity and Nonzero Density of Impurities}

\begin{figure}[h]
\epsfxsize=\columnwidth \centerline{\epsfbox{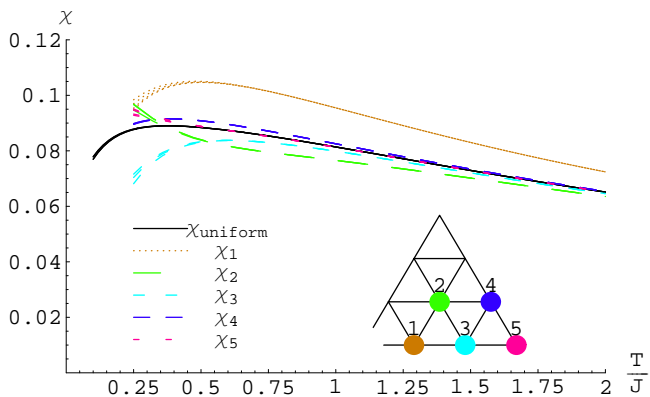}}
\caption{[Color online]
The uniform susceptibility and the local susceptibilities at
the first five inequivalent neighbors of nonmagnetic impurity
(missing site). The four curves for each $\chi_i$ are the Pade
approximants [5,6],[5,7],[6,6],[6,7].
$\chi$ is in units of $(g\mu_B)^2 / J$.
}
\label{fig_local_susc_highT}
\end{figure}

The coefficients of the susceptibility of the first five nearest
neighbors near the impurity are in Table~\ref{tab_local_susc}.
The corresponding local susceptibilities along with the uniform
susceptibility are plotted in Figure~\ref{fig_local_susc_highT}.
We see that the local susceptibility decays to the uniform one in
few lattice spacings.  This is consistent with calculated very short
correlation length in Ref.~\onlinecite{Elstner93}.
The deviation of the near-neighbor local susceptibility from the
uniform value reaches roughly $15\%$ at $T \approx J/3$.

\begin{table*}
\caption{\label{tab_local_susc} Series coefficients $a_n$ of
$\chi = \frac{g^2 \mu_B^2}{T} \sum_{n=0}
\frac{a_n}{4^{n+1} (n+1)\text{!}} (\frac{J}{T})^n$ for susceptibility
$\chi_{\rm uniform}$ of the pure triangular lattice and for the
five closest neighbors $\chi_i$ of nonmagnetic impurity as
indicated in Figure~\ref{fig_local_susc_highT}.
The coefficients for the uniform susceptibility agree with
Ref.~\onlinecite{Elstner93}.}
\begin{tabular}{|r|r|r|r|r|r|r|}
\hline n & $\chi_{\rm uniform}$ & $\chi_1$ & $\chi_2$ & $\chi_3$ & $\chi_4$ & $\chi_5$ \\
\hline
\hline 0 & 1 & 1 & 1 & 1 & 1 & 1 \\
\hline 1 & -12 & -10 & -12 & -12 & -12 & -12 \\
\hline 2 & 144 & 108 & 132 & 138 & 144 & 144 \\
\hline 3 & -1632 & -1248 & -1312 & -1400 & -1560 & -1608 \\
\hline 4 & 18000 & 15840 & 13840 & 13320 & 14880 & 16160 \\
\hline 5 & -254016 & -237024 & -235776 & -213984 & -189168 & -192000 \\
\hline 6 & 5472096 & 4144000 & 5539968 & 5817504 & 5084464 & 4564560 \\
\hline 7 & -109168128 & -73210624 & -93128960 & -109647744 & -123994240 & -118354560 \\
\hline 8 & 818042112 & 1133266176 & 222006528 & 112173696 & 913327488 & 1312247808 \\
\hline 9 & 17982044160 & -18170275840 & 11644656640 & 30128806400 & 38868680960 & 28664414720 \\
\hline 10 & 778741928448 & 581215033344 & 1535178191360 & 1512448745984 & 328581324544 & -104688021504 \\
\hline 11 & -90462554542080 & -21239974981632 & -84715204509696 & -115649955864576 & -118987461639168 & -96786926315520 \\
\hline 12 & 829570427172864 & 215676565092352 & -788032311226368 & -332026092103680 & 2149211723363328 & 2738259718125568 \\
\hline
\end{tabular}
\end{table*}

We would like to see if the observed NMR lines can be explained
from a finite density of such impurities.  The prediction is obtained
by plotting the histogram of susceptibilities.  First we note that at
$T \sim 100 {\rm K} \sim J/3$ the experimental lines are spread by
about $\pm 10\%$ which is roughly equal to the calculated deviation of
the local susceptibility of the nearest neighbors from the uniform
susceptibility.  The experimental curves have a significant weight spread
over this width, and so if the calculated curves are to explain them,
the first few nearest neighbors of impurities would have to form a
sizable fraction of the total number of sites.
Ref.~\onlinecite{Shimizu03} suggests that the system contains about
one Cu impurity in ten thousand which gives about $0.4\%$ fraction for
the nearest neighbors up to $\chi_5$ in Fig.~\ref{fig_local_susc_highT}
and hence it is very far from explaining the experimental lines --
it would predict very sharp histograms.

One possible explanation of this discrepancy is the fact that the
Heisenberg model is not entirely adequate because it would eventually
predict ordering at low temperatures, which is not observed in
experiments, and so there are additional interactions.
Indeed, Ref.~\onlinecite{ringxch} proposed that ring exchanges are
important to stabilize the spin liquid phase.
However, since these interactions are still short-range, the
volume fraction of sites that are affected by vacancies at these
temperatures is still small, so low impurity density cannot fit
the observed data.

A different more likely explanation is that there are more impurities or
more disorder in the system.  One possible source of disorder is from
extended defects such as grain boundaries, and in the next section
we consider the susceptibility near a boundary.
Other possible source is from the ethylene group disorder in the ET
molecules which is thought to be important in a related
$\kappa$-(ET)$_2$Cu[N(CN)$_2$]Br material.
\cite{Soto, Miyagawa, Wolter, Maksimuk}
This can give rise to a large density of point perturbations which are
mild but present everywhere.  Other possible source is disorder in the
(CN)$^{-}$ groups in the insulating anion layers.
\cite{Geiser, Komatsu, Emge}
We currently do not know much about the presence and magnitude of such
perturbations in the spin liquid \ET\ material.

\begin{figure}[h]
\epsfxsize=\columnwidth \centerline{\epsfbox{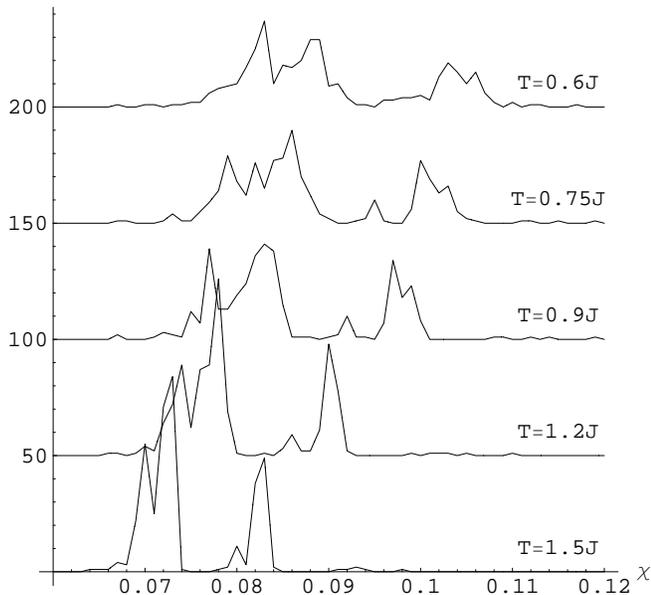}}
\caption{The histogram of local susceptibilities for a $20 \times 20$
triangular lattice sample with 5\% randomly placed vacancies,
obtained from the high temperature series expansion to
11-th order in $J$.  $\chi$ is in units of $(g\mu_B)^2 / J$.}
\label{fig_Triang_HT_susc}
\end{figure}

To simulate a case of a large density of point-like disorder, we
study the system in the presence of a 5\% of missing sites.
Realistic point-like disorder is probably of a different nature,
but vacancies is all we can do in the systematic high temperature
expansion.  However we hope the basic features of the histogram would be
similar.  The result is shown in the Fig.~\ref{fig_Triang_HT_susc}.
Due to a large number of diagram imbeddings we were able only to go
to the 11th order in $J$ (rather than $12$-th as above) on a
$20 \times 20$ lattice and reliably only down to $T \sim 0.5 J$.
The error on the histogram values is roughly $\pm 20\%$.
Crudely, we see two sets of peaks in Fig.~\ref{fig_Triang_HT_susc}:
The one on the right is associated with the nearest neighbors of the
impurities while the one on the left with the rest of the sites.
The finer features are associated with sites at different positions
with respect to several impurities.

As we have already mentioned, the nearest neighbor susceptibility is
different from the the uniform one by about 10\%, which is roughly
similar to the broadening of the NMR line at $T \sim 100-50$~K.
However the temperature dependence of $\chi_1 - \chi_{\rm uniform}$
is also weak, see Fig.~\ref{fig_local_susc_highT}.
In the case with very low density of impurities such as $0.01\%$ Cu,
besides having this value only at very few sites, it would not give
broadening of the lines.
In the case of $5\%$ of impurities, Fig. \ref{fig_Triang_HT_susc},
this corresponds to the distance between the two sets of peaks
not changing significantly with temperature.
On the other hand the (set of) peaks themselves visibly broaden more.
Remembering that here we plot histograms corresponding to the ``ideal''
system (i.e., with only Heisenberg exchanges and vacancies, and
not including other interactions and sources of line widths),
few percent of impurities can indeed produce reasonably broadened
lineshapes at these elevated temperatures.

\subsection{Line Impurity}
\begin{figure}[h]
\epsfxsize=\columnwidth \centerline{\epsfbox{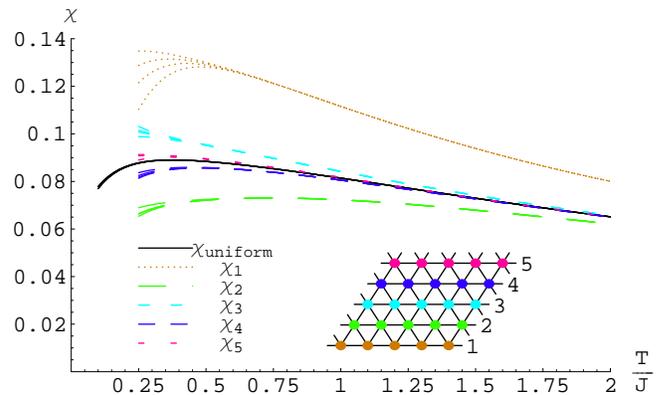}}
\caption{[Color online]
The uniform susceptibility and the local susceptibilities at
the first five inequivalent neighbors of the boundary.  The four
curves for each $\chi$ are the Pade approximants
[5,6],[5,7],[6,6],[6,7].
$\chi$ is in units of $(g\mu_B)^2 / J$.
} \label{fig_line_susc_highT}
\end{figure}

In this subsection we consider pure triangular lattice near a boundary
and calculate susceptibility at various distances from the boundary.
The coefficients of the uniform susceptibility and of the local
susceptibility at the first five closest inequivalent sites are in
Table~\ref{tab_line_susc}.  The corresponding plots are in
Figure~\ref{fig_line_susc_highT}.

\begin{table*}
\caption{\label{tab_line_susc} Series coefficients $a_n$ of
$\chi = \frac{g^2 \mu_B^2}{T} \sum_{n=0}
\frac{a_n}{4^{n+1} (n+1)\text{!}} (\frac{J}{T})^n$ for susceptibility
$\chi_{\rm uniform}$ of the pure triangular lattice and for the five
closest inequivalent neighbors $\chi_i$ to a boundary as indicated in
Figure~\ref{fig_line_susc_highT}.}
\begin{tabular}{|r|r|r|r|r|r|r|}
\hline n & $\chi_{\rm uniform}$ & $\chi_1$ & $\chi_2$ & $\chi_3$ & $\chi_4$ & $\chi_5$ \\
\hline
\hline 0 & 1 & 1 & 1 & 1 & 1 & 1 \\
\hline 1 & -12 & -8 & -12 & -12 & -12 & -12 \\
\hline 2 & 144 & 72 & 120 & 144 & 144 & 144 \\
\hline 3 & -1632 & -752 & -896 & -1440 & -1632 & -1632 \\
\hline 4 & 18000 & 8640 & 5120 & 9040 & 16080 & 18000 \\
\hline 5 & -254016 & -103488 & -108960 & -37248 & -127296 & -230976 \\
\hline 6 & 5472096 & 1497440 & 3972864 & 2808736 & 1342432 & 3429216 \\
\hline 7 & -109168128 & -29967872 & -58795776 & -109978368 & -46504448 & -21990912 \\
\hline 8 & 818042112 & 553745664 & -912840192 & 598482432 & 1331324928 & -895304448 \\
\hline 9 & 17982044160 & -4034237440 & 35460869120 & 74136878080 & -6674631680 & -1900426240 \\
\hline 10 & 778741928448 & -38283289088 & 1453883081728 & -796283040256 & -765905530368 & 2445141614080 \\
\hline 11 & -90462554542080 & -6599243882496 & -88646526167040 & -131119323998208 & 1918538846208 & -58857804742656 \\
\hline 12 & 829570427172864 & 433688769173504 & -1170019148326912 & 3744417183383552 & 1814576120913920 & -3956791382702080 \\
\hline
\end{tabular}
\end{table*}

We see that the ratio of the local susceptibility to the uniform one
is somewhat larger for the first neighbor here than in the single
vacancy case and more importantly the temperature dependence is stronger.
Furthermore, the line impurity is an extended object, so a much
larger number of sites is affected.  We do not know how common such
grain boundaries are in the \ET\ material to make predictions for the
experiment.  However, as in the case with vacancies, we see that
down to $T \sim J/4$ only the first few $\chi_i$ near the boundary
deviate significantly from the bulk value.
Thus to explain the experimental lines with such defects we would
require a large density of them.

\section{Spin Liquid with Fermionic Spinons}
\label{sec:SL}

The \ET\ system does not order down to temperatures as low as $32$~mK,
but has a $J \approx 250$~K. Thus it is a good candidate for spin liquid.
Among SU(2)-invariant spin liquids constructed using fermionic spinons,
the uniform spin liquid has the lowest variational energy in the
relevant model with ring exchanges.\cite{ringxch}
It consists of spinons hopping on triangular lattice with no fluxes
and thus having Fermi surface.
In the full theory, the spinons are coupled to a U(1) gauge field.
\cite{SSLee, LeeNagaosaWen, Ioffe, LeeNagaosa, Polchinski, Altshuler}
This is hard to solve directly.
In order to make progress we solve the problem in the mean field theory
ignoring the gauge field and obtaining a system of free fermions
hopping on the triangular lattice.
We also go beyond mean field by using Gutzwiller projection.

Specifically, in the mean field, we consider free fermions hopping
on the triangular lattice in the presence of a missing site,
a line boundary, a finite density of missing sites,
and also a random distribution of hopping amplitudes.
These are models of non-magnetic disorder in terms of what the
spinons see.

If $\{ \psi_n(i) \}$ is the set of single-particle wavefunctions,
it is easy to show that the local susceptibility at temperature
$T$ is given by
\begin{eqnarray}
\chi(i) &=& \frac{(g\mu_B)^2}{2T} \sum_n
|\psi_n(i)|^2 f(\epsilon_n) (1-f(\epsilon_n)) ~,
\label{eq_chi_freeF}
\end{eqnarray}
where $f$ is the Fermi function $f(\epsilon) =
1/(e^{(\epsilon - \mu)/T}+1)$.  In each model of impurities, we
obtain the wavefunctions and use this formula to obtain the local
susceptibility.
We consider various kinds of disorder in turn.  We present the results,
a very basic discussion, and leave proper discussion of the
possible connection to the \ET\ to a later section.

Below we keep the spinon hopping $t$ fixed and vary the temperature,
and the presented susceptibilities are in units of $(g \mu_B)^2/t$.
In a more systematic calculation, the spinon hopping amplitudes would
need to be found self-consistently for a given spin Hamiltonian and
temperature $T$.
In the clean system, the self-consistent $t$ vanishes above some
temperature of order $J$ (e.g., in the renormalized mean field scheme
this temperature is $0.75 J$).  When $t$ becomes non-zero, this signals
that the system becomes correlated paramagnet, and the spinon
mean field is one attempt to capture the growing local correlations.
Below the onset temperature, the self-consistent $t$ quickly approaches
the zero-temperature value, and it is this regime that we are
describing when keeping $t$ fixed.  We can estimate the spinon hopping
amplitude in the renormalized mean field scheme as
$t = 3 J \la f_i^\dagger f_j \ra \approx 0.5 J$.
The free fermion susceptibility on the half-filled triangular lattice
with such $t \sim 100$K would be $\chi \sim 10 \cdot 10^{-4}$~emu/mol,
which is about a factor of 2-3 larger than the experimental values,
but is reasonable given the serious approximations in such calculations.

We further discuss the self-consistent approach in the case with
random bond disorder in Sec.~\ref{subsec:RandomBonds}.
Before that in the examples below, we introduce the disorder into the
spinon problem by hand, either by simply removing the links to
vacancy sites, or by taking randomly distributed bonds.

\subsection{Point Impurity}
\begin{figure}[h]
\epsfxsize=\columnwidth \centerline{\epsfbox{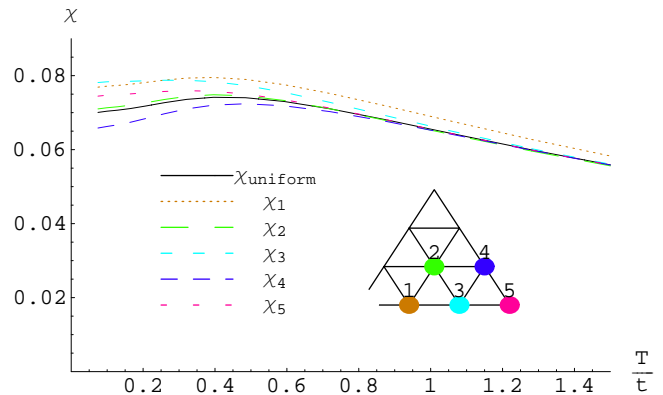}}
\caption{[Color online]
The uniform susceptibility and the local susceptibilities at
the first five inequivalent neighbors of the impurity obtained for
the system of free fermions hopping on the triangular lattice at
half-filling.  The spinon hopping $t$ is kept fixed and the
susceptibilities are in units of $(g\mu_B)^2/t$.
}
\label{fig_local_susc_freeF}
\end{figure}

In this case we find the mean field wavefunctions numerically by exact
diagonalization for system sizes up to $80 \times 80$.
The resulting local susceptibility curves for several neighbors
of the vacancy are in Figure~\ref{fig_local_susc_freeF}.

Precise curves for the susceptibility are hard to obtain at low
temperatures because of the factor $f(1-f)$ in Eq.~(\ref{eq_chi_freeF}),
which becomes increasingly sharp as $T \to 0$.
Thus fewer and fewer states near the chemical potential contribute and
eventually the results are polluted by finite size effects.
Nevertheless, we can go to sufficiently low temperatures with our
system sizes, and the results in Fig.~\ref{fig_local_susc_freeF}
for $T \gtrsim 0.1 t$ essentially represent the infinite-volume limit.
We see that as we lower the temperature, susceptibilities at more and
more neighbors become separated from the uniform susceptibility.

\begin{figure}[h]
\epsfxsize=\columnwidth
\centerline{\epsfbox{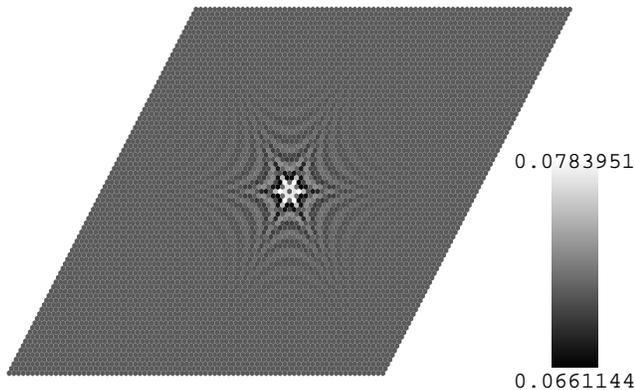}}
\caption{Local susceptibility near impurity at $T = 0.074 t$
obtained from the free fermion calculation for size $80 \times 80$.}
\label{fig_local_susc_freeF_DensityPlot}
\end{figure}

It is interesting to look at the shape of the local susceptibility
as a function of the position.  At a low temperature ($T = 0.054 t$)
this is plotted in Figure~\ref{fig_local_susc_freeF_DensityPlot} as
obtained from the exact lattice calculation.

This distribution converges to a zero temperature distribution.
We can obtain some intuition about the long distance behavior
from a calculation treating the non-magnetic impurity as a perturbation.
The result is
\begin{equation}
\chi_{\rm loc}(r) - \bar\chi = A \frac{\cos(2 k_F r + \phi)}{r}.
\label{eq:chiloc_asymptote}
\end{equation}
Here $k_F$ is the Fermi surface location where the group velocity
points in the observation direction $\vec{r}$, while the phase $\phi$
depends on the impurity type and strength
(just as in the case of Friedel oscillations in metals).
The calculation leading to this result is summarized in
Appendix~\ref{app:chiloc}.

In the present case, the Fermi surface is roughly a circle with
$k_F \approx [2.67,\;\dots,\; 2.72]\, a^{-1}$, where $a$ is the
lattice spacing.
Taking the above expression and plotting it on the lattice gives a
picture looking very similar to
Figure~\ref{fig_local_susc_freeF_DensityPlot}.
One interesting thing to notice is that there is seemingly much longer
wavelength along the $\hat{x}$ direction than $\pi/k_F$.
This simply comes from the fact when we evaluate the
$\cos(2k_F x + \phi)$ on the lattice,
it picks up similar points at different hills of the cosine curve
because the period $\pi/k_F \approx 1.18 a$ is close to one lattice
spacing.

At a finite temperature, the oscillatory power law is cut off
at the characteristic length
\begin{equation}
\xi(T) = \hbar v_F / (2\pi T) ~.
\end{equation}
For the half-filled band on the triangular lattice, the Fermi
velocity $v_F$ does not vary significantly with the direction and
is $v_F \approx [2.86,\;\dots,\; 2.44] t a/\hbar$.
As an example, for $T = 0.1t$ the correlation length is only
$\xi \approx 4a$.

Finally, we would like to know if this distribution, with one
impurity per 10000, can roughly give the observed spectral lines
in the \ET\ material.  The answer is no, and the histograms of
$\chi$ are still negligibly narrow.

\subsection{Line Defect}

\begin{figure}[h]
\epsfxsize=\columnwidth \centerline{\epsfbox{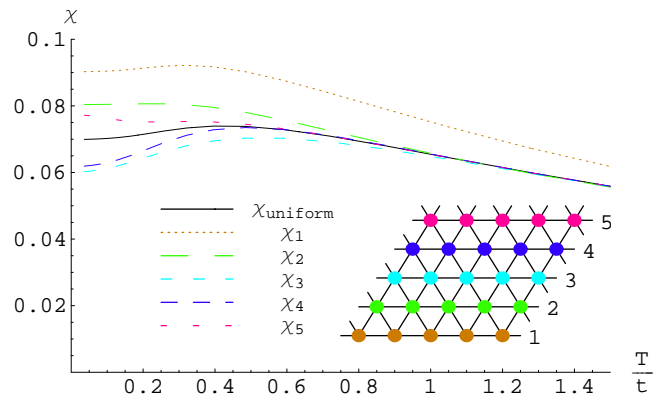}}
\caption{[Color online]
The uniform susceptibility and the local susceptibilities at
the first five inequivalent neighbors of the boundary obtained for
the system of free fermions. $\chi$ is in units of $(g\mu_B)^2/t$.
} \label{fig_line_susc_freeF}
\end{figure}

\begin{figure}[h]
\epsfxsize=\columnwidth \centerline{\epsfbox{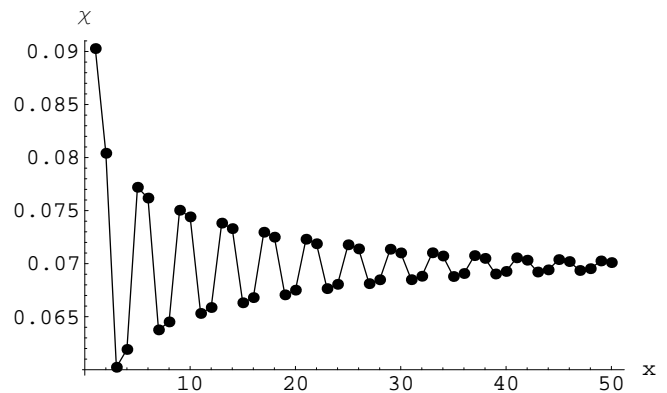}}
\caption{
The local susceptibility as a function of the distance from the boundary
$x$ (line index in Fig.~\ref{fig_line_susc_freeF}) at $T = 0.0364 t$.
Note that the vertical scale does not start at zero and that the
$x \geq 2$ sites lie withing $\pm 15\%$ window around the average.
}
\label{fig_line_susc_freeF_xdep}
\end{figure}

In the clean system with the boundary we can write all wavefunctions
explicitly.  The resulting susceptibilities are shown in
Figure~\ref{fig_line_susc_freeF} for the first five neighbors as a
function of temperature and in Figure~\ref{fig_line_susc_freeF_xdep}
for fixed $T = 0.0364 t$ as a function of the distance from the boundary.

Continuum calculation with a circular Fermi surface predicts at $T=0$
\begin{equation}
\frac{\chi_{\rm loc}(r) - \bar\chi}{\bar \chi} = -J_0(2 k_F r)
\approx -\frac{\cos(2 k_F r - \pi/4)}{\sqrt{\pi k_F r}} ~,
\label{eq:chiloc_boundary_asymptote}
\end{equation}
where $J_0$ is the Bessel function of the first kind and $r$ is the
normal distance from the boundary.
The asymptotic form is valid also for general Fermi surface,
with $k_F$ denoting the momentum where the group velocity is
perpendicular to the boundary.

A finite temperature cuts off the power at the length scale $\xi(T)$:
Roughly, the oscillatory piece is multiplied by
$r/[\xi \sinh(r/\xi)]$.
Indeed a six-parameter function
\mbox{$a_0 + a_1 \cos(a_2 r - a_3) / [r^{a_4} \sinh(a_5 r)]$}
fits the data like that in Fig.~\ref{fig_line_susc_freeF_xdep}
well with $a_2 \approx 2k_F$, $a_4 \approx -0.5$, and
$a_5 \approx 1/\xi$, in agreement with our expectations.
One interesting thing to notice in Fig.~\ref{fig_line_susc_freeF_xdep}
is the apparent period of 4 in terms of the lattice line spacing;
the wavelength in the continuum $\pi/k_F$ is very accurately $4/3$ of
the line spacing, so the apparent period is equal to three wavelengths,
which is an accidental commensuration effect.

\begin{figure}[h]
\epsfxsize=3.5in \centerline{\epsfbox{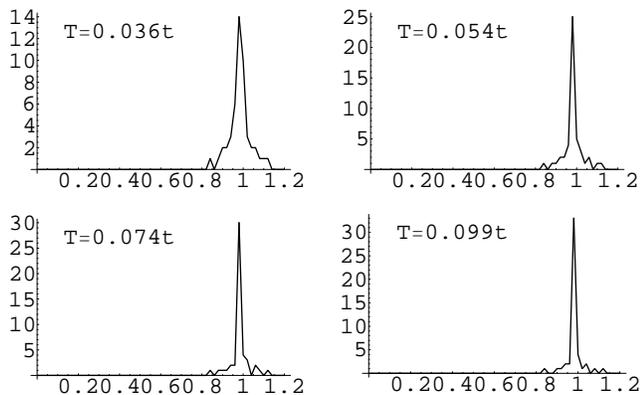}}
\caption{Histogram of the susceptibilities for the system with grain
boundaries separated by 100 lines of sites.}
\label{fig_line_susc_freeF_hist}
\end{figure}

Finally, we look at the histogram of susceptibilities.
This depends on the size of the grain -- in the present model,
the distance between boundaries.  To show an example, we consider
the system as in Fig.~\ref{fig_line_susc_freeF} with boundaries
separated by 100 lines of sites.
The result is in Fig.~\ref{fig_line_susc_freeF_hist}.
The grain boundary can in principle go in many directions or
might not be straight at all.
For the particular orientation that we have chosen, the near
commensuration mentioned above plays a role at low temperatures:
For our grain size, as we lower $T$, the susceptibilities
start to fail to reach the bulk value, and thus we obtain a double peak
in the histogram (one peak from the up hills of the sine curve and the
other from the down hills, cf.~Fig.~\ref{fig_line_susc_freeF_xdep}).
This starts to happen for temperatures just slightly below the
smallest one shown in Fig.~\ref{fig_line_susc_freeF_hist}.

We do not know if this type of disorder is realistic in the \ET.
The chosen separation of 100 spacings between boundaries is a rather
significant disorder: 2\% of the sites are right next to the boundaries
and several times more are in the immediate vicinity.
Still, at temperature above $0.1 t$ the histograms are
very narrow ($\xi$ is still smaller than 5 lattice spacings),
which is why they are shown only below this temperature.
Even at the lowest temperature the linewidth is small,
despite the slow decay of $\chi_{\rm loc}(r)$ away from the boundary.
Thinking about the \ET, it appears that we need more disorder than
this and more spread across the system.

\subsection{5\% of Impurities}
\begin{figure}[h]
\epsfxsize=3.5in \centerline{\epsfbox{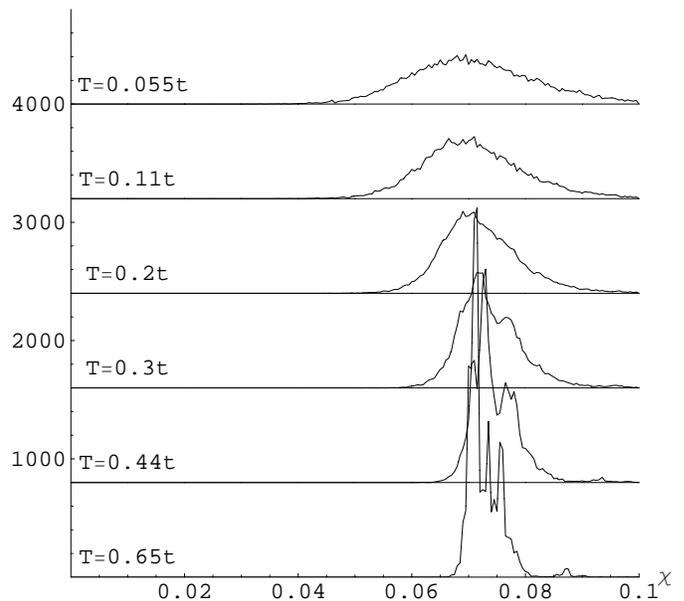}}
\caption{Histogram of the susceptibilities for the system with 5\% of
vacancies.  Note especially that the peaks broaden as we lower the
temperature.  The curves are obtained for $60 \times 60$ samples and are
in the thermodynamic limit for temperatures shown.}
\label{fig_Triang_FF_susc_N60}
\end{figure}

In this section we calculate the susceptibility histogram for the
samples with 5\% of spin vacancies.
It seems unlikely that this type of disorder is present in the \ET
in the form of missing ET dimers.  However, this could be a crude spin
model if the electron charge distribution is inhomogeneous.
There are other likely sources of disorder and this case represents a
situation when the disorder is point-like.
The resulting histograms are in Figure~\ref{fig_Triang_FF_susc_N60}.
We see that peak is quite broad, more like the experimental curves,
and it spreads as we lower the temperature.

\subsection{Random Spinon Hopping Amplitudes}
\label{subsec:RandomBonds}

\begin{figure}[h]
\epsfxsize=3.5in \centerline{\epsfbox{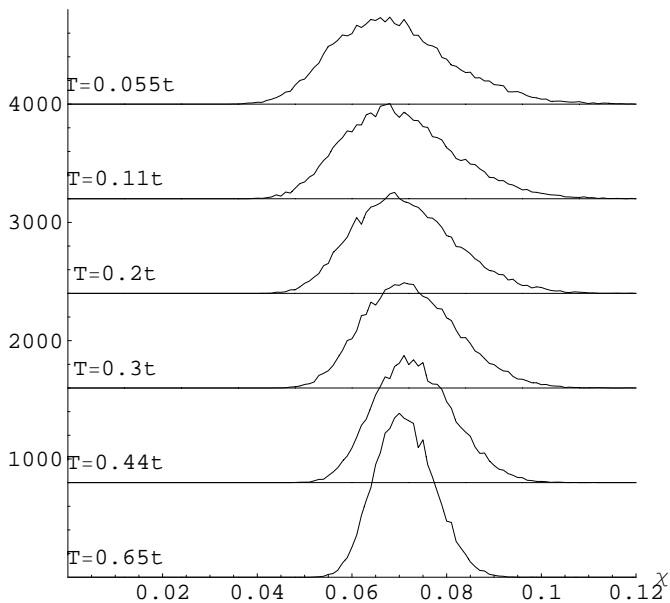}}
\caption{Histogram of the susceptibilities for the system with
random values of bonds with randomness $\pm 20\%$.
Note especially that the peaks do not broaden significantly below
$T \approx 0.44t$.
The curves are obtained for $60 \times 60$ samples and are in the
thermodynamic limit for temperatures shown.}
\label{fig_Triang_FF_bond_dis02_Nx60}
\end{figure}

In this section we take a model of non-magnetic disorder where the
spinon hopping amplitudes are random and uniformly distributed in the
range $[t - \Delta, t + \Delta]$.  The resulting histograms
for $\Delta/t = 0.2$ are in Figure~\ref{fig_Triang_FF_bond_dis02_Nx60}.
Note specifically that the peak of the histogram does not change much
below roughly $T = 0.44 t$ (the peak does not spread).

We also tried to make disorder more point-like and see if this would
cause the peak to spread, as it did for the point-like missing sites
above.  We find that this is indeed the case for the following simple
choice:  Take $95\%$ of bonds to have one value and $5\%$ to have
twice as large value (results not shown).

The fact that point-like disorder spreads the peak upon lowering
temperature can be understood as follows.  As we lower $T$ the
correlation length $\xi(T)$ grows and more and more sites start to
``feel'' the impurities and have susceptibility substantially different
from the bulk value.  This will be happening until the correlation
length becomes somewhat larger than the typical spacing between the
point impurities.  For the samples studied we can go as low as
$T = 0.05t$ (before finite size effects set in), and at this $T$ the
correlation length is roughly ten lattice spacings.
Thus we don't expect the histogram in Fig.~\ref{fig_Triang_FF_susc_N60}
with $5\%$ vacancies to broaden much as the temperature is
lowered further.

So far we took a specific fixed distribution of spinon hopping amplitudes
and calculated susceptibilities.
As we argue below, this is reasonable for a model of non-magnetic
disorder where couplings in the spin Hamiltonian have some randomness
(assuming it is not too strong and the spin liquid state remains stable).
To treat the system more properly, the spinon hopping amplitudes
should be calculated self consistently.
For example, in the Heisenberg model with exchanges $J_{rr'}$,
popular mean field self-consistency conditions read
$t_{rr'}^* = \tilde{J}_{rr'} \la f^\dagger_r f_{r'} \ra$;
the bond expectation value is for one spin species and
$\tilde{J}_{rr'}$ is proportional to $J_{rr'}$,  e.g., in the so-called
renormalized mean field scheme one takes $\tilde{J} = 3J$.
In the clean system, the above self-consistency condition has
non-trivial solution once the temperature is lower than
$T = \tilde{J}/4$.  Rather quickly below this temperature
the spinon hopping becomes large and comparable to the zero-temperature
limit.  However, in this specific treatment, the spin liquid with the
Fermi surface is not a stable solution and other spin liquids perform
better.  Furthermore, as is known from early slave particle studies,
the best states in this mean field are dimerized.  In particular,
if we try to solve the self-consistency equations by iteration starting
from a random initial $t_{rr'}$, these run away towards some
dimerized solutions.

As discussed in the introduction, we expect that there are additional
spin interactions such as ring exchanges that stabilize the zero flux
spin liquid against other spin liquid states and dimerized states.
Ref.~\onlinecite{ringxch} presented a schematic mean field argument
how the ring exchanges achieve this.
The self-consistency condition is modified to
\begin{eqnarray}
t_{rr'}^* &=& \tilde{J}_{rr'} \la f^\dagger_r f_{r'} \ra
+ \sum_{ss'} \tilde{K}_P
\la f^\dagger_r f_s \ra \la f^\dagger_s f_{s'} \ra
\la f^\dagger_{s'} f_{r'} \ra ~,
\label{eq_selfct}
\end{eqnarray}
where $s,s'$ are all the sites so that each expectation value is for
two sites on a bond, and such sum effectively covers all four-site
rhombi $P = [rss'r']$ on the triangular lattice that contain the bond
$r r'$.  The couplings $\tilde{K}_P$ are proportional to the
ring exchanges acting around the rhombi.

Ref.~\onlinecite{ringxch} applied the above scheme to the clean system
at $T=0$ and found that the uniform spin liquid with no fluxes in the
hoppings is a stable solution for $\tilde{K} / \tilde{J} > 9.9$.
Note that the parameters $\tilde{J}$ and $\tilde{K}$ are related to the
microscopic Heisenberg and ring exchanges by disparate numerical factors,
and Ref.~\onlinecite{ringxch} contains more details in what sense such
$\tilde{K} / \tilde{J}$ values are reasonable in the study of the
spin liquid.  Here we mainly use this scheme to have a starting point
where the uniform state is stable in the clean system and see crudely
how the randomness in the microscopic parameters like $J$ translates to
randomness in the spinon hopping amplitudes.

To get some understanding of the self-consistent distribution of
$t_{rr'}$ and its temperature dependence we iterated
Eq.~(\ref{eq_selfct}) until convergence for the following
system parameters.
We took $\tilde{K}_P=15$ everywhere and took a uniformly distributed
$\tilde{J}_{rr'}$ from the interval $[1-\Delta, 1+\Delta]$ for
two values $\Delta = 0.05, 0.2$.
We find that once the nontrivial solutions appear, which happens
quickly below $T \sim 1/4$ (in units of $\tilde{J}$), the distribution of
$t_{rr'}$ is essentially independent of the temperature
and has the same width in relative terms as the distribution of
$\tilde{J}_{rr'}$ but is a bit more rounded.
Calculating histogram of susceptibilities for this distribution gives
roughly the same result as calculating it for the box distribution of
$t_{rr'}$'s of the same relative width as that of $\tilde{J}_{rr'}$.
This provides some justification to the preceding models of disorder
where we simply put randomness into the spinon problem by hand.


\subsection{Gutzwiller Wavefunction Study of the Local Magnetization}
Let us discuss the spin liquid picture beyond the mean field.
One way to proceed is to consider effective gauge theory description
where spinons interact with the emergent gauge field.
One expects that the power laws in
Eqs.~(\ref{eq:chiloc_asymptote},\ref{eq:chiloc_boundary_asymptote})
are modified by the gauge field fluctuations,\cite{Kolezhuk06, Kim03, new_power} but reliable quantitative information is lacking.

In this work we go beyond the mean field by Gutzwiller projection.
In the 1D case, this essentially reproduces exact result\cite{Eggert95}
for the Heisenberg system with a nonmagnetic impurity.
However, in 2D the Gutzwiller projection alone likely does not capture
all important fluctuations in the low-energy theory.\cite{ringxch}
Nevertheless, by working directly with the physical spin variables,
it gives quantitatively more plausible results than the mean field.

Specifically, we consider local magnetization distribution in a
partially polarized state both in the mean field and after the
projection.  We used this approach to study non-magnetic impurities
in a kagome spin liquid in Ref.~\onlinecite{kagome_w_vacancies}
(this reference also contains more discussion on the connection to the
local susceptibilities).

\begin{figure}
\centerline{\includegraphics[width=\columnwidth]{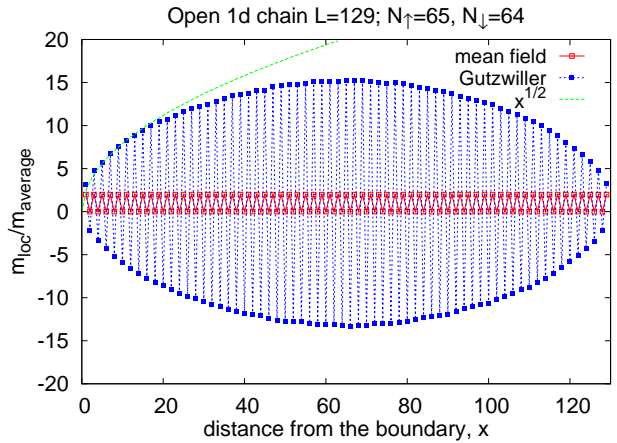}}
\vskip -2mm
\caption{[Color Online]
Local magnetization profile in a 1D chain of length $L=129$ with open
boundaries before and after Gutzwiller projection of a partially
polarized spinon Fermi sea state, here with $N_\up = 65$, $N_\dn = 64$.
The excess up spin occupies orbital $\sin(k_F x) = \sin(\pi x/2)$,
producing a staggered component in $\chi_{\rm loc}$ with a constant
amplitude.  The projection dramatically enhances the staggered component,
which now grows as $\sqrt{x}$ away from the boundary, in agreement with
Ref.~\onlinecite{Eggert95} for the Heisenberg chain.
Such effects can produce broad $\chi_{\rm loc}$ histogram even for
small impurity density in the 1D chain.
We want to contrast this with the 2D case, where we find only a
fixed numerical enhancement.
}
\label{fig:Gutzw1D}
\end{figure}

It is well known that the Gutzwiller-projected Fermi sea is an
excellent trial wavefunction for the 1D spin-1/2 chain, and we test
our approach in this case.  Figure~\ref{fig:Gutzw1D} shows results for a
chain with open boundaries.  In the mean field, the local magnetization
is $\chi_{\rm loc}(x) \sim 1 - \cos(2 k_F x) = 1 - (-1)^x$.
The projection dramatically enhances the staggered component.
In the Heisenberg chain, Eggert and Affleck\cite{Eggert95} predict
that the staggered component in $\chi_{\rm loc}$ grows as $\sqrt{x}$
away from the boundary at $T=0$, and the Gutzwiller-projected state
appears to capture this result in the $m_{\rm loc}$.
This dramatic behavior of the $\chi_{\rm loc}$ near a non-magnetic
impurity has been used to explain broad lines in spin-1/2 chain
compounds even with small density of impurities.\cite{1D_experiments}

\begin{figure}
\centerline{\includegraphics[width=\columnwidth]{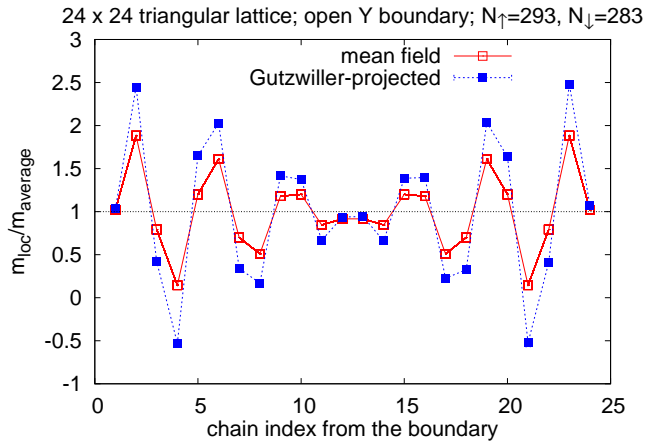}}
\vskip -2mm
\caption{[Color Online]
Local magnetization profile in a triangular lattice with grain
boundaries before and after Gutzwiller projection of a partially
polarized spinon Fermi sea state.
The lattice is constructed of 24 chains of length 24, with periodic
boundary conditions along the chains and open BC in the perpendicular
direction, see the drawing in Fig.~\ref{fig_line_susc_freeF}.
The partial spin polarization is obtained starting from an unpolarized
state and depopulating 5 spin-down orbitals and occupying 5 spin-up
orbitals near the Fermi patch whose normal is perpendicular to the
boundary.
}
\label{fig:Gutzw2D}
\end{figure}

Our initial hope was that the 2D spin liquid, which is also a projected
Fermi sea state and whose full theory shows enhanced spin correlations
at $2k_F$, could similarly produce broad $\chi_{\rm loc}$ histograms
around small density of impurities.
However, it appears that quantitative aspects in 2D are such that
small impurity concentration does not give large line broadening.

Specifically, in the 2D spin liquid case, we studied both a single
impurity and a line boundary and found that $m_{\rm loc} - \bar m$ is
enhanced by projection only by a fixed numerical factor of about two.
Figure~\ref{fig:Gutzw2D} shows representative results in the line
boundary case, where the impurity perturbation is the largest of the
two cases, cf.~Eqs.~(\ref{eq:chiloc_asymptote}) and
(\ref{eq:chiloc_boundary_asymptote}).
The triangular lattice is constructed by stacking 24 chains of length 24,
with periodic boundary conditions along the chains and open boundaries
in the perpendicular direction.  In an effort to bring out more effect,
the excess spin-up population occupies orbitals near the patch where the
Fermi velocity is normal to the boundaries, since it is this $k$-space
region that is responsible for the power law in
Eq.~(\ref{eq:chiloc_boundary_asymptote}).
Note that because of this special population, the mean field amplitude
of oscillations is larger compared to the case with thermal
population of orbitals (irrespective of orientation)
in Fig.~\ref{fig_line_susc_freeF_xdep}.
Nevertheless, the figure shows that the Gutzwiller projection gives
only a fixed enhancement over the mean field by a factor of about two.
Different system sizes and orbital populations do not change this result
qualitatively.  Similar numerical enhancement was observed in our
Kagome study,\cite{kagome_w_vacancies} where we also argued for it
using renormalized mean field thinking.

To conclude, we expect that all our mean field results will experience
a similar numerical enhancement in the $\chi_{\rm loc} - \bar{\chi}$
by the projection, so the histograms will be broader by about a
factor of two.
In particular, we can find negative local susceptibilities, despite the
mean field giving only non-negative $\chi_{\rm loc}$.
But unlike the 1D, we are not able to get small density of impurities to
produce broad lineshapes.
One caveat here is that, as we have already mentioned, the Gutzwiller
projection in 2D does not capture the full gauge theory, and it could be
that the effects of gauge fluctuations are much more dramatic
(e.g., see footnote~\cite{new_power}).  This could happen if the actual
spin liquid phase has much stronger correlations than the mean field
prediction, but at the moment we do not know how to address this better
quantitatively.

\section{Inhomogeneous Knight Shifts in the Metallic Phase}

In Sec.~\ref{sec:SL}, we used free fermions as a mean field for the
spinons.  Assuming the spin liquid is appropriate in the insulator,
the treatment further neglects gauge fluctuations and is a crude
approximation that can change qualitative long-distance behavior.
However, this was best we could do to get some quantitative estimates
of local properties.

In fact, the free fermion analysis applies more readily to the
metallic phase of the \ET.  The fermions are now electrons themselves,
and mean field is reasonable in the Fermi liquid regime.
We use the same formula Eq.~(\ref{eq_chi_freeF}) to calculate the
local susceptibilities.  Also, the analytical results
Eq.~(\ref{eq:chiloc_asymptote}) for the long-distance behavior of
$\chi_{\rm loc}(r)$ away from a single impurity and
Eq.~(\ref{eq:chiloc_boundary_asymptote}) away from a boundary
hold in the metal.

\begin{figure}[h]
\epsfxsize=\columnwidth
\centerline{\epsfbox{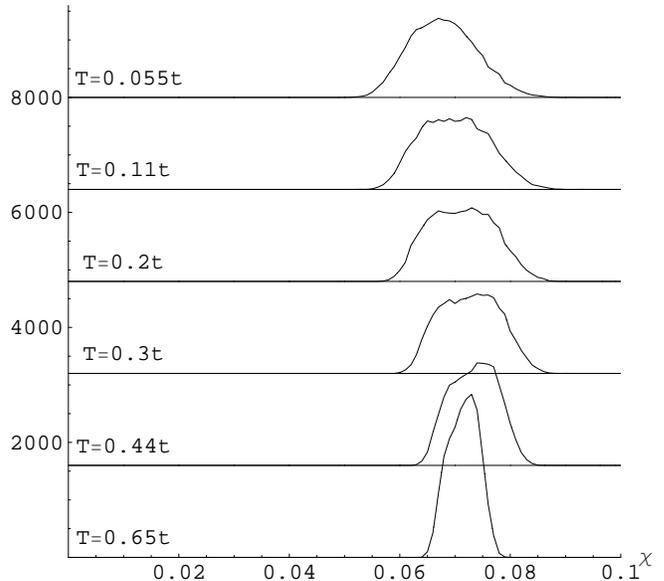}}
\caption{Histogram of the susceptibilities in the metallic phase,
where we model disorder by random on-site potentials uniformly
distributed in an interval $[-W, W]$ with $W = 0.3 t_e$.
The curves are obtained for $60 \times 60$ samples and are in the
thermodynamic limit for temperatures shown.}
\label{fig_Triang_FF_mu_N60_rand03}
\end{figure}

Here we are interested in modelling weak disorder in the metallic phase
and connecting with the $^{13}$C NMR measurements.
In this case, we calculate the histogram of susceptibilities in
the presence of random on-site potentials taken to be uniformly
distributed in an interval $[-W, W]$.
The result for $W = 0.3 t_e$ is in
Figure~\ref{fig_Triang_FF_mu_N60_rand03}.

Here $t_e$ is the electron hopping amplitude and is in excess of 50~meV.
This model of disorder is reasonable in the following sense.
Note that the disorder strength $W$ should be compared to the band width,
which is several times $t_e$, so $W = 0.3 t_e$ is a rather weak disorder.
A crude Born approximation of the electron mean free path due
to elastic scattering in the 2D system gives
$k_F l \approx \hbar^2 v_F^2 / (A \overline{\delta \mu^2})$,
where $A$ is the area per site and $\overline{\delta \mu^2}$ is the
variance of the on-site potential.  For the triangular lattice at
half-filling we have $k_F l \approx 8 t^2 / \overline{\delta \mu^2}$,
which gives $k_F l \sim 250$ for the above disorder.  Direct numerical
estimate of the lifetime of momentum eigenstates in the lattice
system gives comparable number.
Residual resistivities in the \ET\ in the metallic phase imply
$k_F l \gtrsim 50$ and larger, so the disorder that we use is reasonable.

Examining the local susceptibility histograms in
Fig.~\ref{fig_Triang_FF_mu_N60_rand03}, we see that below roughly
$T = 0.5 t_e$ the peak no longer spreads.
This $T$ is comparable to the room temperature, and indeed
the available data in the metallic phase shows little temperature
dependence of the linewidth.\cite{Kawamoto06}
Our model linewidths are reasonable, even though we cannot read off
reliably the inhomogeneous broadening component from the lines
in Ref.~\onlinecite{Kawamoto06}.

In the free fermion case of spinons in the previous section we found
that making the bond disorder more point-like, by changing value of
only a fraction of bonds from the uniform value, the peak of the
histogram spreads as we lower the temperature over a wider temperature
range compared to the case where some randomness is present on all bonds.
In this section we also studied whether similar effect takes place for
electrons with random on-site potentials.  We set $95\%$ of chemical
potentials to zero and $5\%$ to a positive value, chosen to be $0.5t$ in
one run and $2t$ in the other.  We found that indeed the peaks spread
in this case too.  This reinforces the conclusion made above that
more point-like disorder causes the $\chi_{\rm loc}$ histograms
to have stronger temperature dependence of the spread.

\section{Discussion}

We summarize the main results with an eye to connect with the
$^{13}$C NMR experiments in the \ET.

First, our high-temperature series study shows that the local
susceptibility near non-magnetic impurities such as vacancies or
grain boundaries can deviate sizably from the bulk value.
We studied specifically the Heisenberg model with vacancies
and obtained quantitatively accurate results down to $T \sim J/4$.
Even at this low temperature, the local susceptibility is modified
perceptibly only within few lattice spacings of the defects,
cf.~Figs.~\ref{fig_local_susc_highT} and \ref{fig_line_susc_highT}.
On the other hand, the $^{13}$C NMR experiments show broadening
that develops gradually from high temperatures; the linewidth
roughly doubles going from $T \sim J \sim 250$~K down to $50$~K,
at which the FWHM already corresponds to about 20-30\% of the bulk
susceptibility.  In our study, the few close neighbors of the defects
show similar deviation in $\chi_{\rm loc}$.  However, for a small
density of defects, which was our initial assumption, essentially
all sites would be sufficiently away from impurities and we would not
be able to obtain comparably broad histograms.

We studied vacancies as a model of local nonmagnetic disorder,
but we do not expect significant changes for other types of disorder
such as random bonds.  The Heisenberg model is also not entirely
adequate at $T=0$, where multi-spin exchanges likely affect the
ground state, but the role of such terms in a nominally spin model
is less important at higher temperatures.  The Heisenberg model
is thus a reasonable choice at such temperatures and was already
used successfully to understand the bulk spin
susceptibility.\cite{Shimizu03, Zheng}
From the high-temperature study with defects, we are led to ask if
there is more disorder in the system than originally thought.
To be able to reproduce the $T=50$~K lines, we seem to need the
disorder strength comparable to that of few percent vacancy
concentration.  Unfortunately, the way the high-temperature series
work, we cannot study directly more realistic models of disorder
such as random bonds, but our work with vacancies gives a rough idea.

Next, we considered the spin liquid with spinon Fermi surface as a
plausible description of the correlated paramagnet in the
temperature range below 50-100 K and down to few Kelvin.
This is a serious assumption, and even within it we can do quantitative
calculations only in the mean field approximation, supplemented by
Gutzwiller renormalizations.
Proceeding nevertheless, in such spin liquid at low temperatures,
$\chi_{\rm loc}(r) - \bar\chi$ decays with slow power laws away from
defects, cf.~mean field Eqs.~(\ref{eq:chiloc_asymptote})\cite{new_power}
and (\ref{eq:chiloc_boundary_asymptote}),
and many sites can be potentially affected by impurities.
However, quantitative aspects appear to be such that we would not get
visibly broadened histograms even at $T=0$ unless there is a significant
density of impurities.  If we postulate a sizable disorder,
we can get $\chi_{\rm loc}$ histograms comparable to the experimental
ones in the temperature range 50 to 10 K.  We find that the variation
with temperature depends on the type of disorder.
If the disorder is uniformly spread, e.g., all bonds are random in
some range, the peak stops broadening at a relatively high temperature
of about half the overall spinon hopping amplitude.
On the other hand, for a more point-like disorder, the peak keeps
broadening to much lower temperatures.  Thus, if the disorder strength
is fixed, in order to get significant temperature dependence of the
linewidth in the spinon analysis we would need a point-like disorder.

The free fermion mean field applies directly to the metallic side of the
phase diagram of the \ET, which appears for pressures above $0.4$~Gpa.
The mean field fermions are now the electrons themselves.
What is observed\cite{Kawamoto06} are essentially temperature-independent
$^{13}$C NMR lines.  As a reasonable type of disorder in this case
we took a random distribution of the chemical potentials.
Specifically, for a box distribution $[-0.3 t_e, 0.3 t_e]$,
which gives a reasonably large $k_F l \sim 250$, we find
sensible and temperature-independent $\chi_{\rm loc}$ histograms.
On the other hand, making the disorder more point-like, we find that
the histograms broaden as we lower the temperature.  Thus on the
metallic side the NMR lines suggest a uniformly spread disorder.

Returning to the Mott insulator side, it appears that there is
more disorder here than in the same system on the metallic side.
Furthermore, if the disorder is fixed as is reasonable in the metal,
the metallic side suggests it is uniform and not producing the
broadening of the lines, and so this should also be the case on the
Mott insulator side, which contradicts the experiments.
However, let us assume for a moment that the disorder is point-like.
The peak in Fig.~\ref{fig_Triang_FF_susc_N60} produced for $5\%$
vacancies broadens by about a factor of two or three in the range
$T=0.5t$ to $0.05t$.  Using $t \sim 100$~K, in the experiment
this correspond to the range $50$~K to 5~K where we see broadening by
about a factor of three which is thus a reasonable agreement.
As mentioned, we don't expect much broadening below $0.05t$ because the
correlation length is already about ten lattice spacings which is
comparable to the distance between impurities in this example.
However the experiment shows very strong broadening beyond that.
It might be that the origin of the broadening is not disorder and that
our spin liquid picture is not adequate and some other phase is emerging
at low temperatures, perhaps with incipient magnetic
order.\cite{Kolezhuk06, Kim03, Galitski}
There is a way to explain the observed phenomena within spin liquid
picture, but it seems to require that the disorder strength effectively
grows as the temperature is lowered.  Below we speculate on how this
may come about, but more experimental input is needed.

One source of disorder mentioned in the literature for the (ET)-based
organic superconductors is ethylene group disorder.\cite{Soto, Miyagawa}
This was particularly discussed for the $\kappa$-(ET)$_2$Cu[N(CN)$_2$]Br
material, where significant sample to sample variations and cooling rate
dependence were observed.
However, recent studies\cite{Wolter, Maksimuk} suggest that the
amount of such disorder in the $\kappa$-(ET)$_2$Cu[N(CN)$_2$]Br is small
at low temperatures and that perhaps the insulating polymeric layer is
involved.\cite{Wolter}

For the spin liquid material \ET,
literature\cite{Geiser, Komatsu, Kawamoto06} mentions that one of the
(CN)$^{-}$ groups in each unit cell is orientationally disordered
and that such structural disorder can generate random electrostatic
potential throughout the lattice.\cite{Emge}
If such disorder is indeed involved, the question is then why it
does not have comparable pronounced effects on the metallic side.
One possible explanation is good screening of charged impurities in
the metal and progressively weaker and eventually absent screening
in the Mott insulator.
For example, in the metallic phase, the Thomas-Fermi screening length
is small, $\lambda_{TF} = 1/\sqrt{4 \pi e^2 \nu(\epsilon_F)} \sim 1$~\AA,
where we estimated the density of states at the Fermi level by
$\nu(\epsilon_F) = 0.28/(t_e \upsilon)$, $t_e = 50$~meV is the
electron hopping amplitude, and $\upsilon = 850.6$~\AA$^3$ is the
3D volume per triangular lattice site.
On the insulator side, an accurate calculation is harder to make.
Using semiconductor language, we can estimate
$\lambda = \sqrt{k_B T/(4\pi e^2 n)}$, where $n$ is the number
of thermally excited charge carriers.  The resistivity of the insulator
at ambient pressure increases by about four orders of magnitude when the
temperature is decreased from room down to 25 K.  Taking this as an order
of magnitude measure of the change in the density of carriers,
we get $\lambda \sim 10 - 100$~\AA, which is about several lattice
spacings.  So one scenario is that the charged disorder is still well
screened at room temperature but gradually becomes more visible below
100 to 50 K, with the screening essentially absent below about 10 to 5 K.
It might be that this type of disorder is more point like, which further
enhances the broadening of the $\chi_{\rm loc}$ histogram on the
insulator side, but is screened on the metallic side where only weak
and more uniformly spread disorder remains that does not cause
significant histogram broadening at lower temperatures.
In Appendix~\ref{app:randomJ}, we briefly discuss how the charged
disorder in the electronic system may translate to that in the
spin model for the insulator.

At present, we do not know how to estimate the strength of disorder
in the system and whether the above scenario is reasonable.
Unlike the metallic phase, we cannot use the electrical resistivity
as a measure of impurity scattering.
We want to remark though that if the disorder is not too strong so
that the spin-1/2 model with say random couplings is applicable,
our spin liquid construction is still a viable candidate for
the ground state.  As discussed in Sec.~\ref{subsec:RandomBonds},
we can accommodate the bond disorder by adjusting spinon hopping
amplitudes, which are now non-uniform.  From the point of view of
spinons, they are now scattered by this disorder.
Interestingly, if the corresponding elastic mean free path is
sufficiently small, the thermodynamics of the spinon-gauge system
with such diffusive spinons differs from the clean case.
\cite{Ioffe, LeeNagaosa}
For example, the specific heat behaves as
$C_{\rm diffusive} \sim T \log (1/T)$
as opposed to the clean system $C_{\rm clean} \sim T^{2/3}$,
while the thermal conductivity behaves as
$\kappa_{\rm diffusive} \sim T^{1/2}$
as opposed to $\kappa_{\rm clean} \sim T^{1/3}$.
The thermal conductivity measurements could potentially give
an independent estimate of the degree of disorder in the insulator,
while at present the $^{13}$C NMR is our only window onto the
disorder.

\acknowledgments
We thank J.~Alicea, S.~Brown, M.~P.~A.~Fisher, E. Fradkin, T. Senthil,
and R.~R.~P.~Singh for useful discussions and the A.~P.~Sloan Foundation
for financial support (OIM).  We also acknowledge the KITP
``Moments and Multiplets in Mott Materials'' program for hospitality.

\appendix

\section{Perturbative calculation in impurity strength}
\label{app:chiloc}
For a weak non-magnetic perturbation we find, to linear order,
\begin{eqnarray}
\chi_{\rm loc}(q) &=&
\frac{(g\mu_B)^2}{2} u_{\rm imp} \times \\
\times \int_k  &&  \!\!\!\!\!\!\!\!\!
\frac{f(\epsilon_k) [1 - f(\epsilon_k)]
      - f(\epsilon_{k+q}) [1 - f(\epsilon_{k+q})]}
     {T(\epsilon_k - \epsilon_{k+q})} ~,
\end{eqnarray}
where $\epsilon(k)$ describes the clean system band structure,
while $u_{\rm imp}$ is the appropriate impurity matrix element,
for which we neglect any dependence on the momenta.
Analyzing the integral in the $T\to 0$ limit, the $q$-dependence
has singularities on the $2 k_F$ surface of the form
\begin{equation}
-\,\frac{\theta(q > 2k_F)}{2\pi v_F^2 \sqrt{c(q-2k_F)}} ~.
\end{equation}
Here $v_F$ is the Fermi velocity and $c$ the curvature at the
appropriate Fermi surface patch.
Going back to real space, we obtain Eq.~(\ref{eq:chiloc_asymptote})
with $A = -u_{\rm imp}/(2\pi^2 v_F^2 c)$ and $\phi=0$.

The $2k_F$ oscillations of the local susceptibility of free
fermions in the presence of impurities can also be related
non-perturbatively to the calculation of the Friedel oscillations in
the local density,\cite{Kolezhuk06, Kim03}
\begin{eqnarray}
\rho_{\rm loc}(i) &=& \sum_n |\psi_n(i)|^2 f(\epsilon_n) ~; \\
\chi_{\rm loc}(i) &\sim &
\frac{\partial \rho_{\rm loc}(i)}{\partial \mu}
\sim \frac{\partial \rho_{\rm loc}(i)}{\partial k_F} ~.
\end{eqnarray}
This gives one power of $r$ slower decay at large distances in
$\chi_{\rm loc}(r)$ than in $\rho_{\rm loc}(r)$ away from defects.

\section{From Random Electron Potential to Random Heisenberg J}
\label{app:randomJ}

Here we discuss schematically how disorder in the microscopic
electronic model translates to disorder in the effective spin
description of the insulator.
For an illustration, consider a Hubbard model
\begin{equation}
H = U \sum_r n_{r\up} n_{r\dn}
- \sum_{rr'} t_{rr'} c_{r\sigma}^\dagger c_{r'\sigma}
+ \sum_r v_r c_{r\sigma}^\dagger c_{r\sigma} ~,
\end{equation}
with hopping amplitudes $t_{rr'}$ and on-site potentials $v_r$.
Large $U$ opens a charge gap.  To leading order in $1/U$, we obtain
spin-1/2 model with Heisenberg couplings
\begin{equation}
J_{rr'} = \frac{4 t_{rr'}^2}
               {U \left[1 - \frac{(v_r - v_{r'})^2}{U^2} \right]} ~.
\end{equation}
Thus, if we have either random hopping amplitudes or random potentials,
the Heisenberg couplings are also random.

The randomness in the electron hopping amplitudes translates directly
to randomness in the spin exchanges -- the latter is even larger in
relative terms.
On the other hand, if the disorder is in the potentials and if these are
much smaller than $U$, this randomness is effectively renormalized down.
This is natural since the variables in the spin model are charge-neutral
and should be oblivious to such potential disorder.
Still, the random potentials are seen via virtual excitations and
induce non-magnetic randomness in the spin system.

If we need to include multi-spin exchanges so as to stabilize
the spin liquid, there will be some randomness in these as well.
Despite the randomness in the effective spin model and irrespective
of its electronic origin, as long as it is not too strong, we can
proceed with the spin liquid construction but now the spinon hopping
amplitudes become non-uniform.
This was discussed in Sec.~\ref{subsec:RandomBonds}.

Finally, if the random potentials become comparable to the Hubbard $U$
and lead to significant variation in the electron density, the very
spin model thinking can become inappropriate.


\end{document}